\documentclass[10pt]{iopart}

\usepackage{graphicx}
\usepackage{epsfig}
\usepackage{bm}
\usepackage{color}
\usepackage{float}
\usepackage{dcolumn}
\usepackage{amssymb}
\usepackage{verbatim}

\def\1{\'{\i}}
\graphicspath{ {images/} }

\begin{document}

\title{Error analysis of nuclear forces and effective interactions}

\author{R. Navarro P\'erez, J.E. Amaro and E. Ruiz Arriola} 
\address{Departamento de
  F\'{\i}sica At\'omica, Molecular y Nuclear \\ and Instituto Carlos I
  de F{\'\i}sica Te\'orica y Computacional \\ Universidad de Granada,
  E-18071 Granada, Spain.}
\eads{\mailto{rnavarrop@ugr.es}, \mailto{amaro@ugr.es}, \mailto{earriola@ugr.es}}

\date{\today}

\begin{abstract} 
\rule{0ex}{3ex} The Nucleon-Nucleon interaction is the starting point
for {\it ab initio} Nuclear Structure and Nuclear reactions
calculations. Those are effectively carried out via effective interactions
fitting scattering data up to a maximal center of mass
momentum. However, NN interactions are subjected to statistical and
systematic uncertainties which are expected to propagate and have
some impact on the predictive power and accuracy of theoretical
calculations, regardless on the numerical accuracy of the method used
to solve the many body problem. We stress the necessary conditions
required for a correct and self-consistent statistical interpretation
of the discrepancies between theory and experiment which enable a
subsequent statistical error propagation and correlation analysis. We
comprehensively discuss an stringent and recently proposed
tail-sensitive normality test and provide a simple recipe to implement
it. As an application, we analyze the deduced uncertainties and
correlations of effective interactions in terms of Moshinsky-Skyrme
parameters and effective field theory counterterms as derived from the
bare NN potential containing One-Pion-Exchange and Chiral
Two-Pion-Exchange interactions inferred from scattering data.
\end{abstract}


\pacs{03.65.Nk,11.10.Gh,13.75.Cs,21.30.Fe,21.45.+v}
\vspace{2pc}
\noindent{\it Keywords}: NN interaction, One Pion Exchange, Chiral Two
Pion Exchange, Statistical Analysis, Effective Interactions \\ 


\section{Introduction}

Quantum Chromodynamics (QCD) is the underlying $SU(N_c)$ theory of
$N_c$ coloured quarks and $N_c^2-1$ gluons ($N_c=3$) which is expected
to explain all known hadronic phenomena from the structure of the pion
to the binding and interactions among atomic nuclei. For light quarks
and disregarding electroweak interactions one expects this to be done
in terms of just two scales: the pion weak decay constant $f_\pi$ and
the pion mass $m_\pi$.  The complexity of the problem for finite
nuclei, neutron and nuclear matter has distorted the very usage of the
term {\it ab initio} calculations~\cite{Meissner:2014cba}; instead of
referring to the solution from QCD in terms of quarks, gluons and
their interactions, it is meant the solution of the many-nucleon
problem from the phenomenological knowledge of the two and/or three
nucleon systems. This distinguishes phenomenological but realistic
nucleon-nucleon interactions as the starting point for any {\it ab
  initio} nuclear structure and nuclear reactions calculations, a
long-standing framework whose parentage to QCD may be sought either by
direct lattice QCD calculations~\cite{Aoki:2011ep} or indirect QCD
NN-features such as chiral symmetry within an Effective Field Theory
(EFT)
approach~\cite{Weinberg:1990rz,Ordonez:1993tn,Kaiser:1997mw,Rentmeester:1999vw,Entem:2003ft,Epelbaum:2004fk,Machleidt:2011zz}
or large $N_c$ scaling of the NN
potential~\cite{Muther:1987sr,Kaplan:1996rk,Banerjee:2001js,
  CalleCordon:2008cz,CalleCordon:2009ps}. Any approach, even when
assisted by these fundamental constraints, must obviously be validated
by confronting NN-forces to scattering data and deuteron properties
before pursuing studies in heavier nuclei. This requires, in particular,
the selection of a CM momentum range $p \le \Lambda$ fixing a de
Broglie resolution scale $\Delta r = \hbar /\Lambda$ and which can be
suitably tuned to target the relevant effective coarse grained
interaction~\cite{NavarroPerez:2011fm} operating in a finite nucleus.

For {\it finite} number of fitting data with {\it finite} experimental
accuracy, neither the theoretical phenomenological interaction nor the
experimental data are free from inconsistencies and systematic bias.
This issue becomes more acute when the number of data {\it and} model
parameters is large. There is of course no infallible method to discard
some subset of inconsistent data out of a finite number of experiments
with finite precision, nor a subset of phenomenological interactions
on the basis of a quantitative disagreement between theory and
experiment. However, statistical methods allow to {\it check} within a
prescribed probabilistic confidence level when both the most likely
theory and a given set of experiments differ by just statistical
fluctuations. This most often happens when the total set of data is
reduced after some, presumably inconsistent, data are discarded and
the model representing the data contains a sufficiently large number
of independent
parameters~\cite{Stoks:1993tb,Stoks:1994wp,Wiringa:1994wb,Machleidt:2000ge,Gross:2008ps}.

Data fitting by least squares minimization is a broadly used technique
in phenomenological approaches in nuclear physics and more generally
in all experimental sciences with satisfactory results. Much has been
said on this topic and we refer to a recent and readable overview of
the subject~\cite{Dobaczewski:2014jga}. In this paper we concentrate
on the applicability conditions and assumptions underlying the
procedure. While our remarks have general validity we will discuss the
particular case of nuclear forces, and more specifically NN scattering
below pion production
threshold~\cite{Perez:2013mwa,Perez:2013jpa,Perez:2013oba,Perez:2014yla}.
Our presentation relies heavily on these works where further details
may be found. 

The goodness of a least squares fit is decided on the value of
$\chi^2/\nu$ in probabilistic terms. Thus, one estimates the
probability of discarding a theoretical model assuming it was a true
one, a so-called type I error, see Sect.~\ref{sec:norm-test} below. A
key assumption which, if fulfilled, provides a sound mathematical
basis for confidence level estimates and statistical error propagation
is the normality of residuals measuring the discrepancy between the
fitting model and the fitted data.  For a finite number of data one
can only give an answer to this question in probabilistic terms, and
of course the conditions to answer in the affirmative are increasingly
stringent with the number of fitted data. We want to stress that
despite the elementary textbook character of this normality
test~\cite{evans2004probability}, this part of the least-squares
fitting is too often overlooked, perhaps because it can only be made
{\it a posteriori} once the minimization has been finished and
potentially invalidating the analysis~\footnote{Many canned routines
  for fitting (for example MINUIT~\cite{james2004minuit},
  POUNDERS~\cite{Kortelainen:2010hv}) of current use do not include
  this useful test at the moment of writing this contribution.}. Our
purpose here is to give simple and straightforward tools enabling such
an analysis and hopefully to popularize it. More specifically, we
quickly overview the normality testing and provide a convenient method
which focuses on the selection process of mutually inconsistent data.


As mentioned, phenomenological interactions are obtained by fitting
scattering data up to a maximum CM momentum $\Lambda$, and thus there
is an inherited $\Lambda$ dependence. We address here a remark
in~\cite{Dobaczewski:2014jga} {\sl "We are dealing almost everywhere
  with effective theories justified in terms of general arguments, but
  whose parameters are basically unknown and often cannot be deduced
  from ab-initio modeling"}.  In fact, coarse graining the interaction
down to the relevant scale allows to carry the identification of
relevant parameters. This is the idea underlying the $V_{\rm low k}$
approach~\cite{Bogner:2001gq,Bogner:2009bt} which universalizes the
interaction and can be visualized from a Wilsonian renormalization
point of view by means of the Block-Diagonal Similarity
Renormalization group (BD-SRG) equations~\cite{Anderson:2008mu}.  As
it was shown in Ref.~\cite{Arriola:2010hj} for the NN problem and also
in the case of diatomic systems~\cite{Arriola:2010tu}, effective
couplings can indeed be obtained just from scattering properties and
for $\Lambda \lesssim 1/a$, with $a$ the range of the interaction. The
complementarity of this method with the full BD-SRG approach for
$\Lambda \lesssim 200 {\rm MeV}$ has been
checked~\cite{Arriola:2013era}. A different approach, directly fitting
parameters up to a maximum energy yields very similar results for
these $\Lambda$-values~\cite{NavarroPerez:2013iwa}. Here we provide
error estimates and correlations on the parameters of the effective
interactions when fitting up to CM momentum $p \lesssim \Lambda=400
{\rm MeV}$.

One of the applications of the normality test discussed previously is
that statistical error propagation including correlations may
confidently be undertaken.  In a series of papers we have addressed
this issue concerning the NN interaction itself, nuclear matrix
elements and possible implications for the accuracy of binding
energies~\cite{NavarroPerez:2012vr,Perez:2012kt,Amaro:2013zka}.  An
error analysis of the empirical mass formula was first discussed in
Ref.~\cite{Toivanen:2008im}. The Predictive power and theoretical
uncertainties of mathematical modelling for nuclear physics has been
studied in Ref.~\cite{dudek2013predictive} within mean field theory
approaches. Here we will focus on the expected statistical
uncertainties of effective interactions in the NN sector.

\section{Statistical framework}

\subsection{Self-consistent least squares fit}

Given $N$ experimental observations $O_i^{\rm exp}$ with estimated
errors $\Delta O_i^{\rm exp}$ the traditional figure of merit for a
theory with $P$ parameters ${\bf p}=(p_1, \dots, p_P)$, predicting
theoretical observable values ${\cal O}_i({\bf p})$, corresponds to
minimizing the least squares sum
\begin{eqnarray}
\min_{\bf p} \chi^2({\bf p}) = 
\min_{\bf p} 
\sum_{i=1}^{N} 
\left( 
\frac{O_i^{\rm exp}-O_i({\bf p})}{\Delta O_i^{\rm exp}}
\right)^2 
\equiv 
\chi^2 ({\bf p}_0) \, . 
\end{eqnarray}
Then the most likely theory
parameters are ${\bf p}_0$, 
and the most likely theoretical prediction is
 ${\cal O}_i^{\rm th} \equiv O_i({\bf
  p}_0).$ 
The residuals at ${\bf p}_0$ are defined by
\begin{eqnarray}
R_i  = \frac{O_i^{\rm exp}-O_i({\bf p}_0)}{\Delta O_i^{\rm exp}}.
\end{eqnarray}
If $R_i$ obey a standardized normal distribution, i.e., 
the probability of obtaining residuals in the interval $[a,b]$ is 
\begin{eqnarray}
P( a \le R_i \le b) = \int_a^b N(x) dx \, , \qquad N(x)=\frac{e^{-x^2/2}}{\sqrt{2\pi}} \, , 
\label{eq:gauss-r_i}
\end{eqnarray}
then $\chi^2 ({\bf p_0}) $ follows a $\chi^2$-distribution with $\nu=
N-P$ degrees of freedom. A confidence region
 in the parameter space around ${\bf p}_0$ may be defined 
by the set of values  $\Delta {\bf p}$ such that
$
\chi^2 ({\bf p}_0 + \Delta {\bf p} ) - \chi^2 ({\bf p}_0 ) 
= \Delta {\bf p}^T {\cal E}^{-1} \, \Delta {\bf p} \le 1, 
$
where ${\cal E}_{ij}$ is the error matrix. The correlation ${\cal
  C}_{ij}$ matrix is defined as
\begin{eqnarray}
{\cal C}_{ij}= \frac{{\cal E}_{ij}}{\sqrt{{\cal E}_{ii}{\cal E}_{jj}}} \, . 
\label{eq:corrmat}
\end{eqnarray}
Thus, for any function of the parameters $F({\bf p})$ one has that to
$1\sigma$ confidence level
\begin{eqnarray}
F({\bf p}) = F({\bf p}_0) \pm \Delta F \, , \qquad (\Delta F)^2 
=   \nabla_{\bf p} F^T \, {\cal E} \, \nabla_{\bf p} F |_{{\bf p}={\bf p}_0} \, . 
\end{eqnarray}
Being $\chi^2$ a positive function the minimum always exists; the
relevant issue concerns the requirement for a meaningful
interpretation. As mentioned, the least squares method rests on the
major assumption that discrepancies between theory and experiment are
random normal variables. Thus, we have to decide whether {\it any} of
the $R_i$ individually fulfills Eq.~(\ref{eq:gauss-r_i}), a question
which could only be answered by repeating the measurement. Instead,
one decides for the ensemble $ ( R_1\, , \dots \, , R_N) $ as a whole,
a question which for a sample with finite size $N$ can only be
answered in probabilistic terms and {\it a posteriori}, i.e. after the
least squares fit has been obtained. In the case of scattering
experiments we deal with a Poissonian statistics counting, which for
moderately large number of counts becomes a gaussian distribution, so
we do expect in the absence of systematic errors in the measurements
and the theory the $R_i$ behaving as normalized gaussians. 
When this happens we have a self-consistent fit and we can write
\begin{eqnarray}
{\cal O}_i = {\cal O}_i^{\rm th} + \xi_i \Delta {\cal O}_i \, , 
\end{eqnarray}
where $\xi_i$ are standardized normal uncorrelated variables, and
$\Delta {\cal O}_i$ are the experimental uncertainties. This formula
has the virtue of simulating a synthetic set of individual and
independent measurements. This has been recently exploited for a study
of errors in fitting parameters of the NN force~\cite{Perez:2014jsa}
and applied to estimate the statistical error in the theoretical
binding energy of the triton~\cite{Perez:2014laa}.

\subsection{Normality tests}
\label{sec:norm-test}

For a set on $N$ empirical data $X_i$ the \emph{null hypothesis} $H_0$
of a normality test is that the data follows a standard normal
distribution i.e. $H_0:X_i \sim N(0,1)$. Correspondingly, the
\emph{alternative hypothesis} $H_1$ states that the empirical data
follows a distribution $F_1$ different from $N(0,1)$. The test
consists of quantitatively assessing if certain discrepancies between the
empirical data distribution and $N(0,1)$ are large enough to confidently
reject $H_0$; these discrepancies are quantified using a \emph{test
statistic} $T$. The decision to reject (or not) is made by comparing
the \emph{observed} test statistic $T_{\rm obs}$, which is calculated using
the empirical data $X_i$, against the distribution of $T$ for random samples
of $N(0,1)$ with size $N$. Each normality test has its own definition of $T$
and in some cases the distribution of $T$ under the null hypothesis is 
known analytically. When comparing $T_{\rm obs}$ to the distribution of 
$T$ a \emph{significance level}  $\alpha$ is arbitrarily chosen, this
determines a critical value $T_c$ and $H_0$ is rejected if $T_{\rm obs}$
is greater (or smaller, depending on the distribution of $T$) than $T_c$.
Common choices for $\alpha$ are $0.01$ and $0.05$. Another relevant quantity 
commonly quoted in normality tests is the $p$-value and corresponds to the
smallest significance level at which $H_0$ would be rejected. A small $p$-value indicates clear
deviations from  normality, whereas a large $p$-value indicates
that no statistically significant discrepancies where found. 

As with any test, one of two type of errors can occur when a decision is made.
One is giving a false negative, also known as type I error, which
consists on rejecting $H_0$ when the data do follow the normal distribution;
the other is a false positive, or type II error, which is made when $H_0$
is not rejected and the data follows a different distribution. The probability
of a type I error is given by the value $\alpha$ and therefore can be 
arbitrarily fixed. With this in mind one would like $\alpha$ to be small
enough to avoid a false negative; but not so small that almost any set of 
empirical data passes the test, rendering the test useless. The probability
of a type II error (or false positive rate) is denoted by $\beta$ and is
directly related to the power of the test, given by $1-\beta$. The statistical
power is an intrinsic property of the test that is more appropriately analyzed 
\emph{a priori} by applying the test to random samples drawn from 
non normal distributions (in this case one knows that the null hypothesis
is false) and seeing how often $H_0$ is not rejected. The power of a test
usually increases with the sample size $N$, and therefore a proper power analysis 
is needed to know how large should $N$ be to avoid giving a false positive.

\subsection{Finite sample statistical fluctuations of residuals}

If $X$ is a continuous random variable following a given probability
distribution $\rho(x)$, the Cumulative
Distribution Function (CDF) is defined as 
\begin{equation}
 \label{eq:generalCDF}
 p(x) \equiv P(X < x ) = \int_{-\infty}^x \rho (t) dt
\end{equation}
Clearly $p(-\infty)=0$ and $p(\infty)=1$. For example, the CDF of the
standardized normal distribution is given
by
\begin{equation}
 \label{eq:NormalCDF}
 \Phi(x) = \int_{-\infty}^x N(t) dt = \frac{1}{2} \left[  {\rm erf} \left(\frac{x}{\sqrt{2}}\right) + 1 \right] \, .
\end{equation}
If we generate $N$ independent data $x_i$ following a certain
distribution $\rho(x)$ we define the empirical CDF (ECDF) as the fraction of
data that is smaller then a certain value $x$ i.e.
\begin{equation}
 \label{eq:EmpiricalCDF}
 S_N(x) = \frac{1}{N}\sum_{i=1}^N\theta(x-x_i).
\end{equation}
Of course the specific ECDF depends on the on the particular sample
$(x_1, \dots, x_N)$; if we take $M$ different extractions $(x_1^{(i)},
\dots, x_N^{(i)})$, $i=1,\ldots,M$, then 
 $M$ different ECDF's $S_N^{(i)}(x)$ will be
generated. Thus, for a function of these $N$ random variables $O(x_1,
\dots, x_N)$ we can define as usual the expected value as the arithmetic mean 
in the limit of $M\to \infty$ 
\begin{eqnarray}
 E[O] 
&\equiv&
 \lim_{M \to \infty} \frac{1}{M}\sum_{i=1}^M O (x_1^{(i)} , \dots , x_N^{(i)} )
\nonumber\\
&=& 
\int_{-\infty}^\infty dx_1 \rho(x_1) \dots \int_{-\infty}^\infty dx_N \rho(x_N) O(x_1, \dots, x_N) \, . 
 \label{eq:meanO}
\end{eqnarray}
One can thus compute the probability density of having a certain value $S$ of
$S_N(x)$. Actually, since by definition we have $S_N(x)=n_N(x)/N$ with $n_N(x)$
an integer number $0 \le n_N(x) \le N $, we may instead compute the
expected probability of finding a {\it given}
value $n_N(x)=m$ as 
\begin{eqnarray}
P_{N,x} (m) &=& E ( \delta_{m,n_N(x)} ) = E \left( \frac1{2\pi}
\int_0^{2\pi} d\varphi \, e^{i \varphi (m-n_N(x))} \right) \nonumber \\ &=&
\frac1{2\pi}
\int_0^{2\pi} d\varphi \,
e^{i \varphi m } \left[ p(x) + e^{-i \varphi} (1-p(x)
  )\right]^N \, , 
\label{eq:meanCDF}
\end{eqnarray}
which after expanding the binomial and computing the integral becomes
a binomial distribution
\begin{equation}
P_{N,x} (m)  = \left( \begin{array}{c} N \\ m
     \end{array} 
\right) p(x)^ m \left[1-p(x)\right]^{N-m} \, . 
\label{eq:binomial}
\end{equation}
The expected value and the variance are thus  
\begin{equation}
E[m] = p(x) \qquad (\Delta m)^2 = \frac1{N} p(x)(1-p(x)) \, . 
\label{eq:mean-bin}
\end{equation}
Matters become simpler for large samples, $N \gg 1$, the binomial 
becomes a normal distribution 
\begin{equation}
P_{N,x} (m) \to 
\frac{1}{ \sqrt{2\pi}\Delta m(x) } 
\exp \left[-\frac12 \left( \frac{m/N-p(x)}{\Delta m(x)}\right)^2  \right] \, , 
\label{eq:bin-gauss}
\end{equation}
so that we may write with $n\sigma$ confidence level 
\begin{equation}
S_N (x) = p(x) \pm  \frac{n}{\sqrt{N}} \sqrt{p(x)(1-p(x))} \, . 
\label{eq:meanCDF2}
\end{equation}

\subsection{Rotated Quantile-quantile plot}

The mapping
\begin{eqnarray}
z= p(x)= \int_{-\infty}^x \rho(t) dt \, , 
\end{eqnarray}
transforms any distribution into the uniform one $U[0,1]$ since $dz =
\rho(x) dx$ and $0 < z < 1$. For a set of $N$ uniformly distributed
discrete points $z_n=n/(N+1)$ with $n=1, \dots, N$ we define their
corresponding $\rho$-theoretical values are defined as
\begin{equation}
\frac{n}{N+1}= p(x_n^{\rm th}) = \int_{-\infty}^{x_n^{\rm th}} dt \rho(t)  \, , 
\label{eq:meanCDF3}
\end{equation}
which fulfill $x_1^{\rm th} < \dots < x_N^{\rm th}$ since the mapping
is monotonous $dz/dx = \rho(x) > 0$.  The empirical {\it
  quantile-quantile} (QQ)-plot of the ranked set of points $ x_1^{\rm
  exp}< \dots< x_N^{\rm exp} $ is defined by plotting the points
$(x_n^{\rm exp}, x_n^{\rm th})$ for $n=1, \dots, N$. One expects that
{\it if} the empirical points follow the distribution $\rho(x)$ then
in the limit of $N \to \infty$ the QQ-plot should become a straight
line $\lim_{N \to \infty} (x_n^{\rm exp}- x_n^{\rm th})=0$. For finite
$N$ there are, however, finite size fluctuations and therefore a
departure from the straight line.

How large can be $\Delta x_{n}^{\rm exp} = x_{n}^{\rm exp}- x_n^{\rm
  th}$ be before we suspect that $x_n^{\rm exp}$ do not follow the
distribution $\rho(x)$?. For large $N$ we can use the normal
distribution $N(x)$, Eq.~(\ref{eq:NormalCDF}),
and fluctuations are more clearly displayed in terms of the {\it rotated}
QQ plot. At the $1\sigma$ confidence level 
\begin{equation}
\Delta x_n = \pm 
\sqrt{ \frac{ p(x_n)(1-p(x_n)) }{ N } } 
\frac{1}{ N(x_n) } \, , 
\label{eq:delxi}
\end{equation}
where we have explicitly used that $x_{n}^{\rm exp}- x_n^{\rm th}=
{\cal O}(1/\sqrt{N})$ to estimate the r.h.s. This is a quick method
providing point-wise normality test bands when $N \gg 1 $ in a rotated
QQ-plot. In Ref.~\cite{Perez:2014yla} we have applied a variety of
traditional methods including the Pearson test, the Kolmogorov-Smirnov
(KS) test, the moments method and more recent Tail sensitive (TS)
test~\cite{Aldor2013} which we describe in the next subsection in a
simplified way.  We illustrate the situation for the rotated-QQ-plots
in Fig.~\ref{fig:qq-plots} (see discussion below) with the band
suggested by Eq.~(\ref{eq:delxi}) when we take as the empirical
$x_i$'s the residuals corresponding to the NN analysis.

\subsection{Tail sensitive test}


The idea behind the tail sensitive (TS) test comes from the
QQ-plot. The Monte Carlo scheme proposed in  Ref.~\cite{Aldor2013} was pursued in
Ref.~\cite{Perez:2014yla}. Here we provide a shortcut to avoid the
large number of samples necessary for the MonteCarlo simulation.  The
recipe for checking normality, or any probability
distribution, $\rho(x)$, 
for a
empirical sample $x_1 < \dots < x_N$ of size $N$, consists of a
few steps
\begin{enumerate}
\item Decide what is the significance level $\alpha$ (typically 0.05
  or 0.01), which is the
  probability of type I error, i.e. giving a false negative. 
\item Determine the critical test statistic $T_c$, which  
for $N> 50$ corresponds to 
\begin{equation}
T_c = \frac{a}{\sqrt{N}}+b 
\end{equation}
where $a$ and $b$  depend on $\alpha$ 
and can be looked up in table~\ref{tab:TSparameters}. For $N<50$ one can look at table~\ref{tab:TStestcritical}. 
\item Transform the empirical data $x_1 < \dots < x_N $ to the new 
 data $z_1 < \dots < z_N $ through the mapping
\begin{equation}
z_i = \int_{-\infty}^{x_i} dt \rho(t)
\end{equation}
\item Compute the test statistic
$T= 2 
\min_i 
\left\{ \min \left[ B_{i,N+1-i}(z_i) , 1 - B_{i,N+1-i}(z_i) \right] 
\right\},
$ 
where the cumulative distribution function corresponds to the
regularized incomplete Beta-function and is defined as
\begin{equation}
B_{i,N+1-i} (z)=  \sum_{j=i}^{N}
\left( \begin{array}{c} N \\ j
     \end{array} 
\right) z^j (1-z)^{N-j}
\end{equation}
\item Compare the observed $T$ with the critical theoretical value
  $T_c$. If $T \le T_c $ the assumption that the empirical data $ x_1,
  \dots , x_n $ were drawn from the probability distribution $\rho(x)$
  can be rejected with a confidence level of
  $100(1-\alpha)\%$. If $T \ge T_c$ there are no
  statistically significant reasons to reject the assumption. This
  is usually  expressed in short saying that the sample
  follows the distribution $\rho(x)$ with 
  $100(1-\alpha)\%$ confidence level.
\end{enumerate}

\subsection{The Birge factor}

\begin{figure}[htbp]
\begin{center}
\epsfig{figure=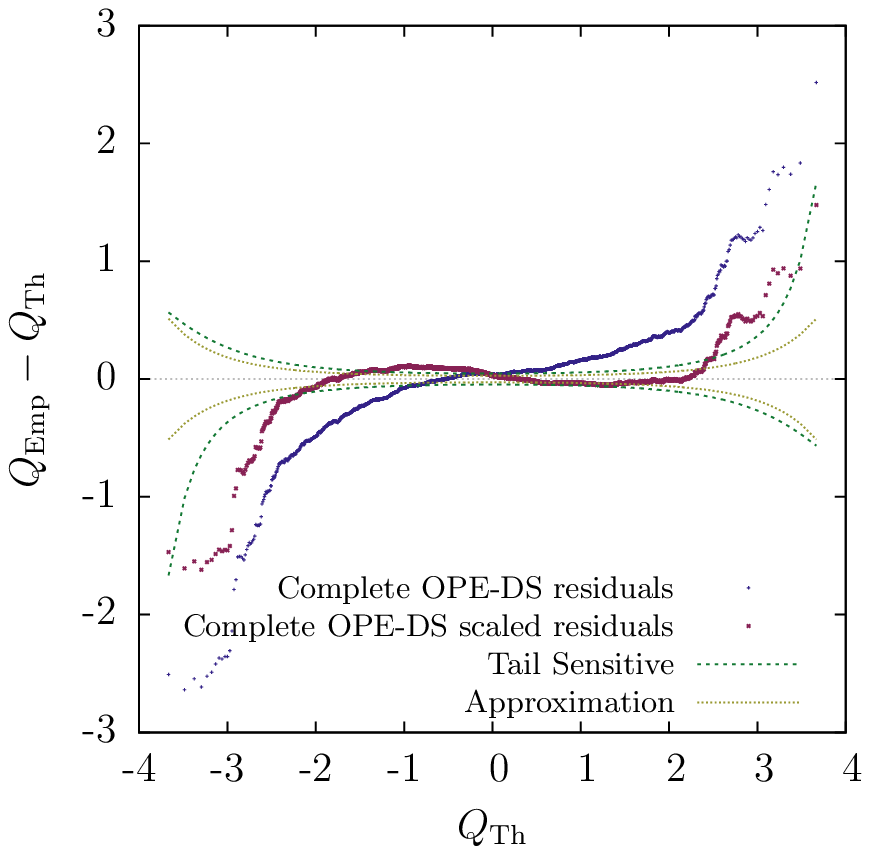,width=6cm,height=6cm} 
\epsfig{figure=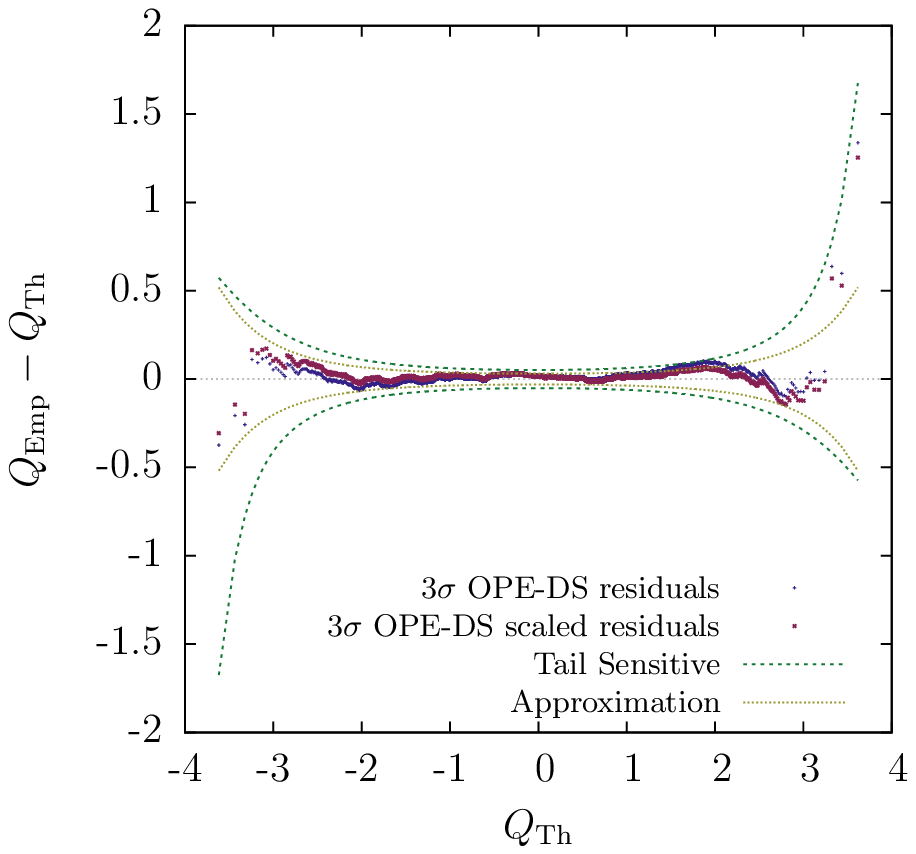,width=6cm,height=6cm} 
\epsfig{figure=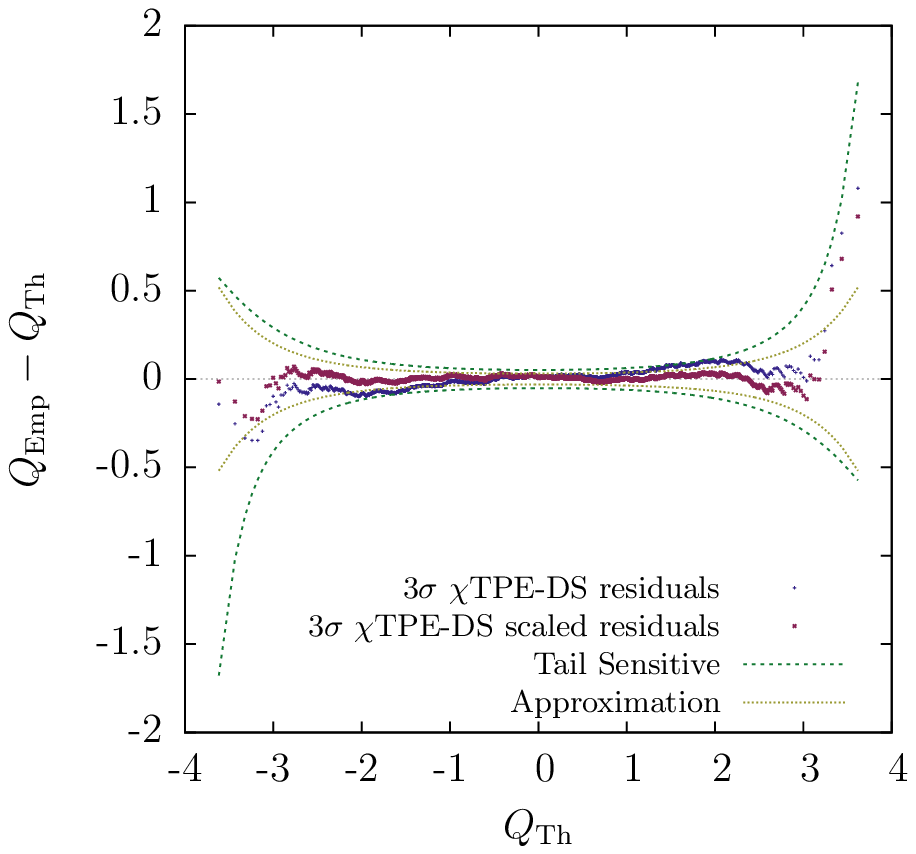,width=6cm}  
\end{center}
\caption{ (Color online) Rotated QQ-plots with point-wise,
  Eq.~(\ref{eq:delxi}), and TS $95\%$-confidence bands compared with
  data residuals before and after Birge rescaling of errors. 
  Top-left panel: Full database with $N=8125$ data residuals with
  OPE-DS~\cite{Perez:2013jpa}.  
  Top-right panel: Selected database with
  $N=6173$ date residuals with OPE-DS~\cite{Perez:2013jpa}.  
  Lower panel: Selected database with $N=6173$ date residuals with
  $\chi$TPE-DS~\cite{Perez:2013oba}.  }
\label{fig:qq-plots}
\end{figure}

When $\chi_{\rm min}^2/\nu $ is outside the expected $1\sigma$
confidence interval, $1\pm \sqrt{2/\nu}$, an artificial and global
rescaling enlarging the errors is often
recommended~\cite{Dobaczewski:2014jga}. From this point of view the
long struggle of selecting a NN database for a high quality
phenomenological
potential~\cite{Stoks:1993tb,Stoks:1994wp,Wiringa:1994wb,Machleidt:2000ge,Gross:2008ps,Perez:2013mwa,Perez:2013jpa,Perez:2013oba,Perez:2014yla}
would be meaningless. However, as shown in Ref.~\cite{Perez:2014yla}
this is reasonable {\it once} the normality of residuals has been
tested. We illustrate this rescaling in Fig.~\ref{fig:qq-plots} for
the NN analysis on the light of the normality TS test at the
$95\%$-confidence level. For the complete $N=8125$ database analyzed
with OPE-DS (complete references to published data are provided in
Ref.~\cite{Perez:2013jpa}), a lack of normality is evident both before
($\chi^2/\nu=1.4$) and after rescaling. However, the $N=6173$
3$\sigma$-mutually self-consistent selected database complies with
normality before ($\chi^2/\nu=1.04$) and after rescaling when analyzed
with OPE-DS.  This same database only complies with normality after
rescaling when analyzed with $\chi$TPE-DS (before rescaling
$\chi^2/\nu=1.07$)~\cite{Perez:2013oba}. This shows that rescaling is
only justified {\it provided} normality of residuals has been
achieved~\cite{Perez:2014yla}.

\section{Effective interactions and their uncertainties}

\subsection{Motivation}

An important result from the statistical correlation analysis is that
we may actually learn what is the most convenient scheme to search for
the optimized interaction. Obviously, by diagonalizing the correlation
matrix ${\cal C}_{ij}$ we may determine what are the linear
combinations of fitting parameters which behave independently along
the minimization path. However, this is inconvenient as does not
precisely correspond to clear picture in the parameter space.  In what
follows we study this issue from the point of view of effective
interactions as applied to the OPE-DS~\cite{Perez:2013jpa} and
$\chi$TPE-DS~\cite{Perez:2013oba} interactions. We are using the same
$3\sigma$ self-consistent database of $N=6713$ data of
Ref.~\cite{Perez:2013jpa} which as described above passes
satisfactorily the TS normality test~\cite{Perez:2014yla}.

\subsection{Moshinsky-Skyrme parameters}

\begin{figure}[ht]
\begin{center}
\epsfig{figure=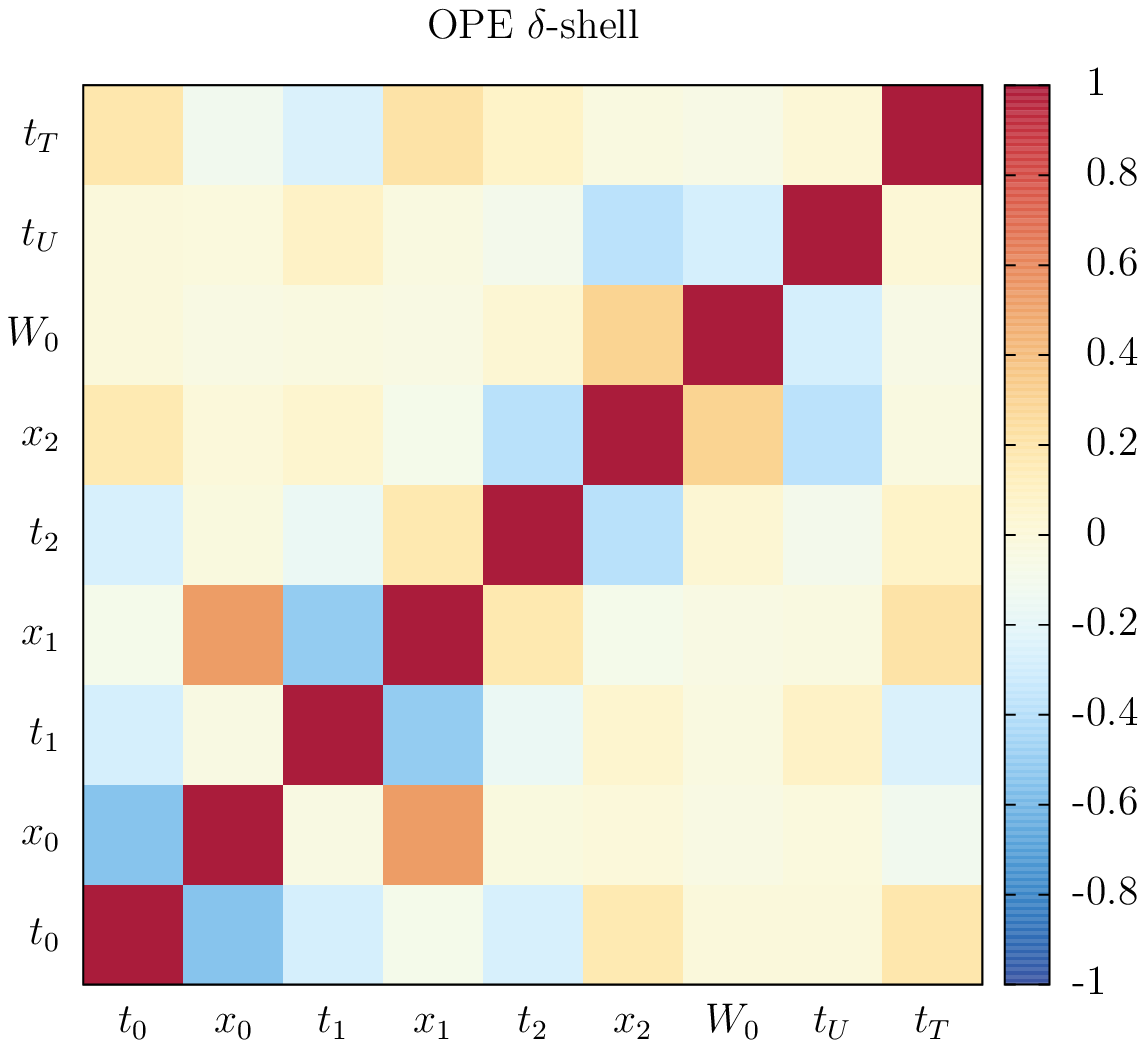,width=6cm}  
\epsfig{figure=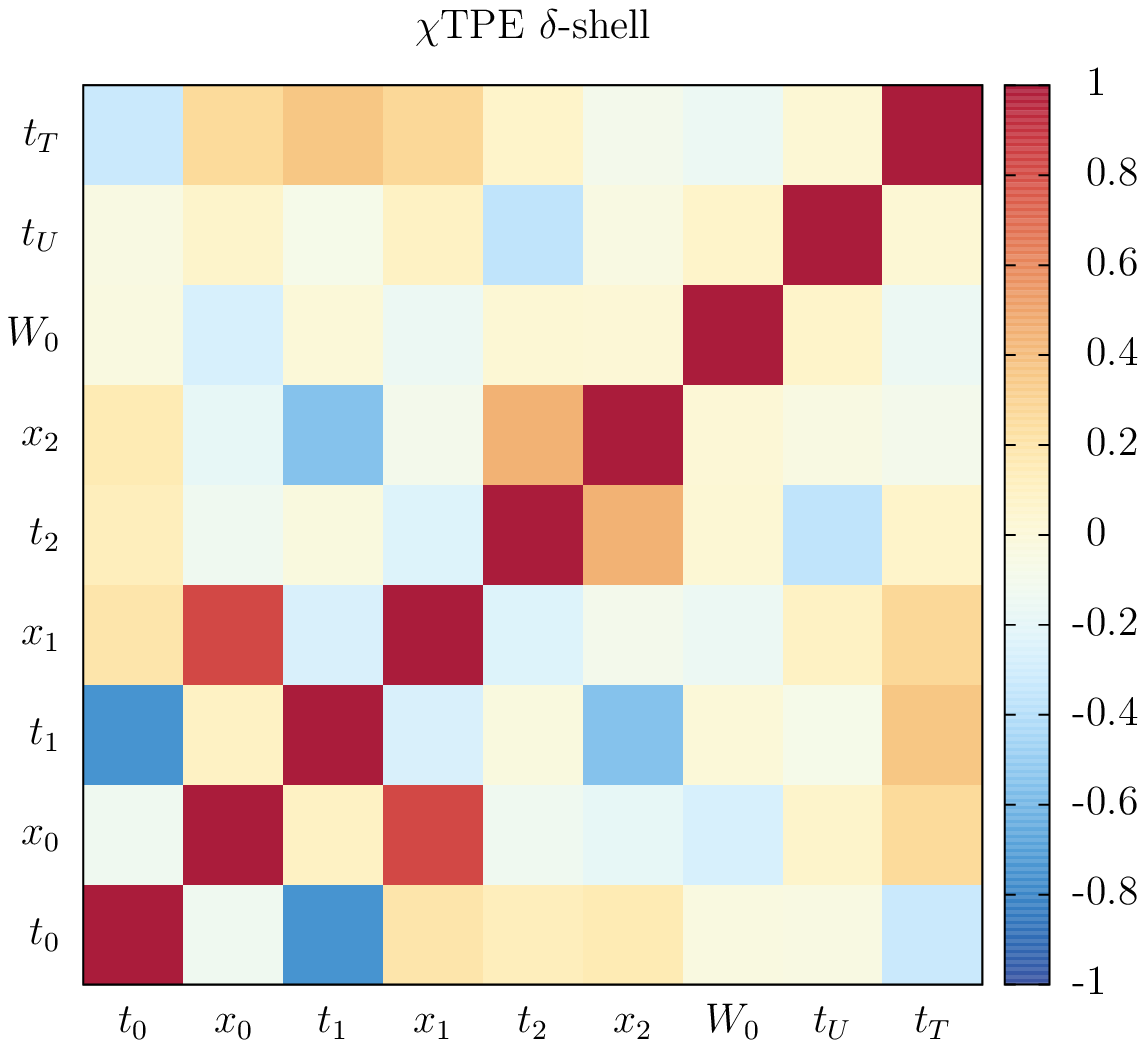,width=6cm}  
\end{center}
\caption{ (Color online) Correlation matrix ${\cal C}_{ij}$,
  Eq.~(\ref{eq:corrmat}), for the 9 Skyrme parameters, see
  Eq.~(\ref{eq:skyrme2}). We show the OPE-DS (left panel) and
  the $\chi$TPE-DS (right panel) potentials. We grade from $100\%$
  correlation, ${\cal C}_{ij}=1$ (dark red), $0\%$ correlation, ${\cal
    C}_{ij}=0$ (light yellow) and $100\%$ anti-correlation, ${\cal
    C}_{ij}=-1$ (dark blue).}
\label{fig:corr-skyrme}
\end{figure}

Effective interactions in Nuclear Physics were proposed by
Moshinsky~\cite{Moshinsky195819} and Skyrme~\cite{Skyrme:1959zz}.  As
compared to {\it ab initio} calculations, the nuclear many body wave
function has a much simpler structure since short range correlations
play a marginal role allowing for a fruitful implementation of mean
field Hartree-Fock
calculations~\cite{Vautherin:1971aw,Chabanat:1997qh,Bender:2003jk}.
At the two body level the effective interaction of
Moshinsky~\cite{Moshinsky195819} and Skyrme~\cite{Skyrme:1959zz} can
be written as a pseudo-potential in the form
\begin{eqnarray} 
&& V_\Lambda ({\bf    p}',{\bf p}) 
=
 \int d^3 x e^{-i {\bf x}\cdot ({\bf p'}-{\bf p})}  \hat V({\bf x} ) 
 \nonumber \\ &&=  t_0 (1 + x_0 P_\sigma ) + \frac{t_1}2(1 + x_1
  P_\sigma ) ({\bf p}'^2 + {\bf p}^2) \nonumber \\ 
&&+  
 t_2 (1 + x_2
  P_\sigma ) {\bf p}' \cdot {\bf p} + 2 i W_0 {\bf S} \cdot({\bf p}'
  \wedge {\bf p}) \\ &&+ 
\frac{t_T}2 \left[ \sigma_1 \cdot {\bf p}
  \, \sigma_2 \cdot {\bf p}+ \sigma_1 \cdot {\bf p'} \, \sigma_2
  \cdot {\bf p'} - \frac13 \sigma_1 \, \cdot 
\sigma_2 ({\bf p'}^2+  {\bf p}^2)
\right] \nonumber \\  &&+
\frac{t_U}2 \left[ \sigma_1 \cdot {\bf p}
  \, \sigma_2 \cdot {\bf p}'+ \sigma_1 \cdot {\bf p'} \, \sigma_2
  \cdot {\bf p} - \frac23 \sigma_1 \, \cdot 
\sigma_2 {\bf p'}\cdot  {\bf p}
\right]  
+ {\cal O} (p^4) \nonumber 
\label{eq:skyrme2}
\end{eqnarray} 
where $P_\sigma = (1+ \sigma_1 \cdot \sigma_2)/2$ is the spin exchange
operator with $P_\sigma=-1$ for spin singlet ($S=0$), and $P_\sigma=1$
for spin triplet ($S=1$) states. The cut-off $\Lambda$ specifies the
maximal CM momentum scale, and $\Delta r=\hbar/\Lambda$ the de Broglie
resolution.  The scale dependence of the parameters was determined in
Ref.~\cite{Arriola:2010hj} just from NN threshold properties such as
scattering lengths, effective ranges and volumes without explicitly
taking into account the finite range of the NN interaction, a
procedure which is well justified for $\Lambda \lesssim 200 {\rm MeV}$
when the Similarity Renormalization Group is
invoked~\cite{Arriola:2013era}.  One can likewise take $\Lambda$ as
the maximal fitting CM momentum~\cite{NavarroPerez:2013iwa}. Here we
will take the fixed value $\Lambda=400 {\rm MeV}$ and exemplify our
results for both OPE-DS~\cite{Perez:2013jpa} and
$\chi$TPE-DS~\cite{Perez:2013oba}.  The expansion contains just those
9 parameters which can be determined from two-body data
alone~\cite{Arriola:2010hj}.  Using the formulas
from~\cite{NavarroPerez:2013iwa} expressing the Skyrme parameters as
volume integrals we get the numerical results of
Table~\ref{tab:SK}. We separate the contributions stemming from the
inner region $r < r_c$ containing just delta-shell interactions and
the outer region $r>r_c$ containing the pion exchange potential tail.
It is noteworthy the numerical agreement between the full integrals of
both potentials. In Fig.~\ref{fig:corr-skyrme} we show the
correlation matrices corresponding to the inner parts. In both cases
correlations are small. 

As mentioned and demonstrated in~\cite{NavarroPerez:2013iwa} the
numerical values of the parameters depend on the resolution
scale. Therefore, a direct numerical comparison of our values with
mean field approaches should not be taken as a measure of agreement;
actually we expect that a suitable $\Lambda~\sim 100-200 {\rm MeV}$
would produce closer numbers~\cite{Arriola:2010hj}.

\begin{table}[ht]
 \caption{\label{tab:SK} Moshinsky-Skyrme parameters 
for the renormalization scale $\Lambda=400 {\rm MeV}$. 
  We separate the contribution from the delta-shells short
   range parameters (corresponding to $r < r_c$) and the potential
   tail (corresponding to $r> r_c$) both for OPE-DS ($r_c=3 {\rm fm}$)~\cite{Perez:2013jpa} and $\chi$TPE-DS ($r_c=1.8 {\rm fm}$)~\cite{Perez:2013oba}. Units
   are: $t_0$ in ${\rm MeV} {\rm fm}^3$, $t_1,t_2,W_0,t_U,t_T$ in ${\rm MeV} {\rm fm}^5$,
and $x_0,x_1,x_2$ are dimensionless.}
\begin{center}
 \begin{tabular}{c c c c  c c c }
\hline
                       & \multicolumn{3}{c}{OPE-DS($r_c=3 {\rm fm}$)} &  \multicolumn{3}{c}{$\chi$TPE-DS($r_c=1.8 {\rm fm}$)} \\[-5pt] 
                     &   $ r < r_c $ & $ r > r_c $ & Full &     $ r < r_c $ & $ r > r_c $ & Full \\              
\hline                     
$t_0$ & -490.6(64)   & -136.2       & -626.8(64)  &  -170.9(70)   & -358.7(32)   & -529.6(53)  \\[-5pt]
$x_0$ &   -0.49(2)   &    0.032     &   -0.38(2)  &    -1.55(7)   &   -0.0934(8) &   -0.56(1)  \\[-5pt]
$t_1$ &  357.7(30)   &  590.4       &  948.1(30)  &   114.7(29)   &  798.8(21)   &  913.6(22)  \\[-5pt]
$x_1$ &   -0.218(9)  &    0.055     &   -0.048(3) &    -0.53(2)   &   -0.00855(2)&   -0.074(3) \\[-5pt]
$t_2$ &  407.5(56)   & 2055.1       & 2462.6(56)  &   230.6(31)   & 2259.4(46)   & 2490.0(39)  \\[-5pt]
$x_2$ &   -1.118(4)  &   -0.8190    &   -0.8686(6)&    -0.71(1)   &   -0.892(1)  &   -0.8750(8)\\[-5pt]
$W_0$ &  107.7(4)    &    0         &   107.7(4)  &    96.1(4)    &    4.7       &  100.8(3)   \\[-5pt]
$t_U$ &  392.9(12)   &  885.7       &  1278.6(12) &   127.8(6)    & 1132.5(7)    & 1260.3(5)   \\[-5pt]
$t_T$ &-1204.4(87)   &-3016.5       &-4220.9(87)  &  -457.7(31)   &-3835.1(21)   &-4292.8(23)  \\
\hline 
 \end{tabular}
\end{center}
\end{table}

\subsection{Counterterms and their uncertainties}

\begin{figure}[htbp]
\begin{center}
\epsfig{figure=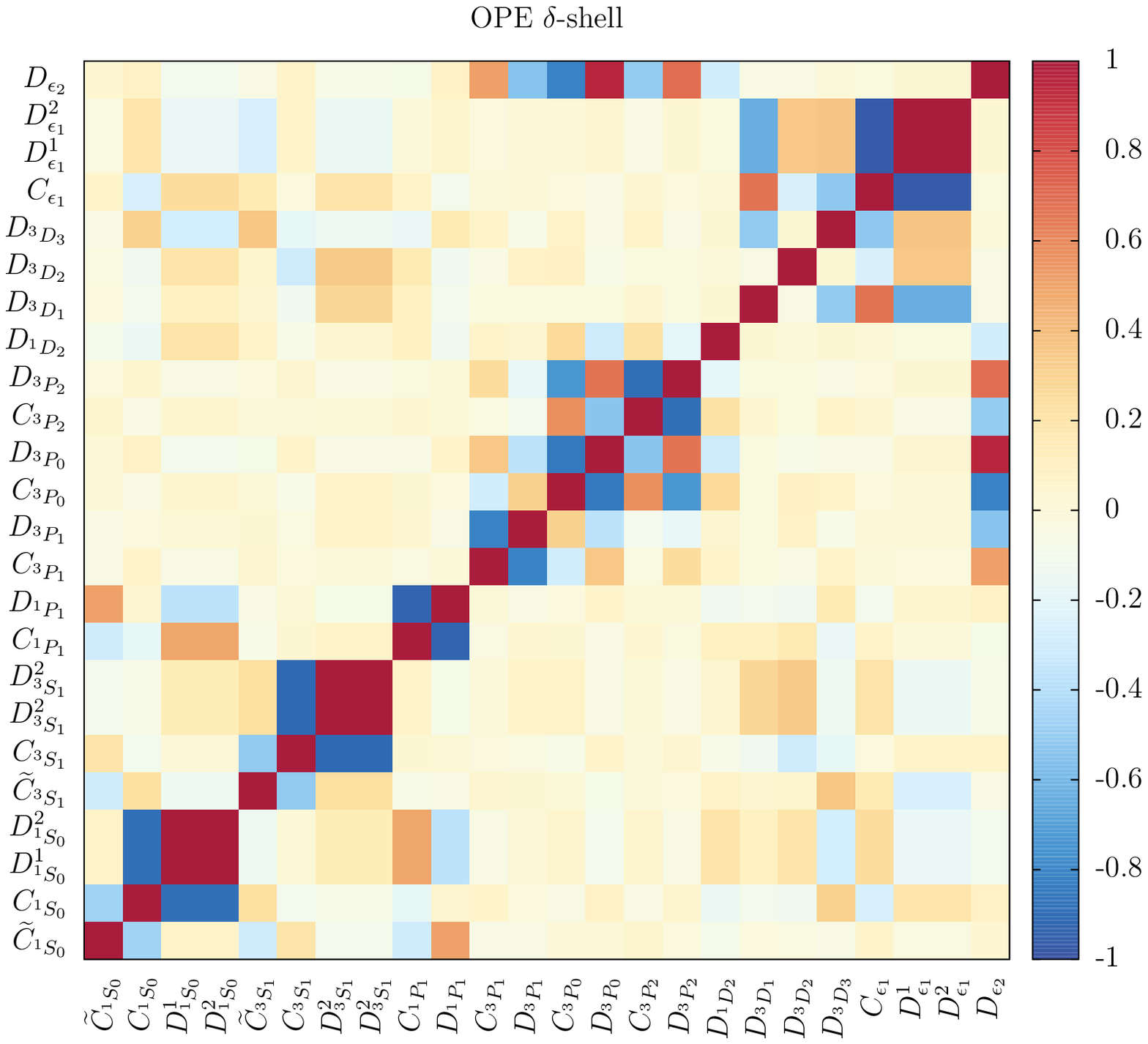,width=6cm} 
\epsfig{figure=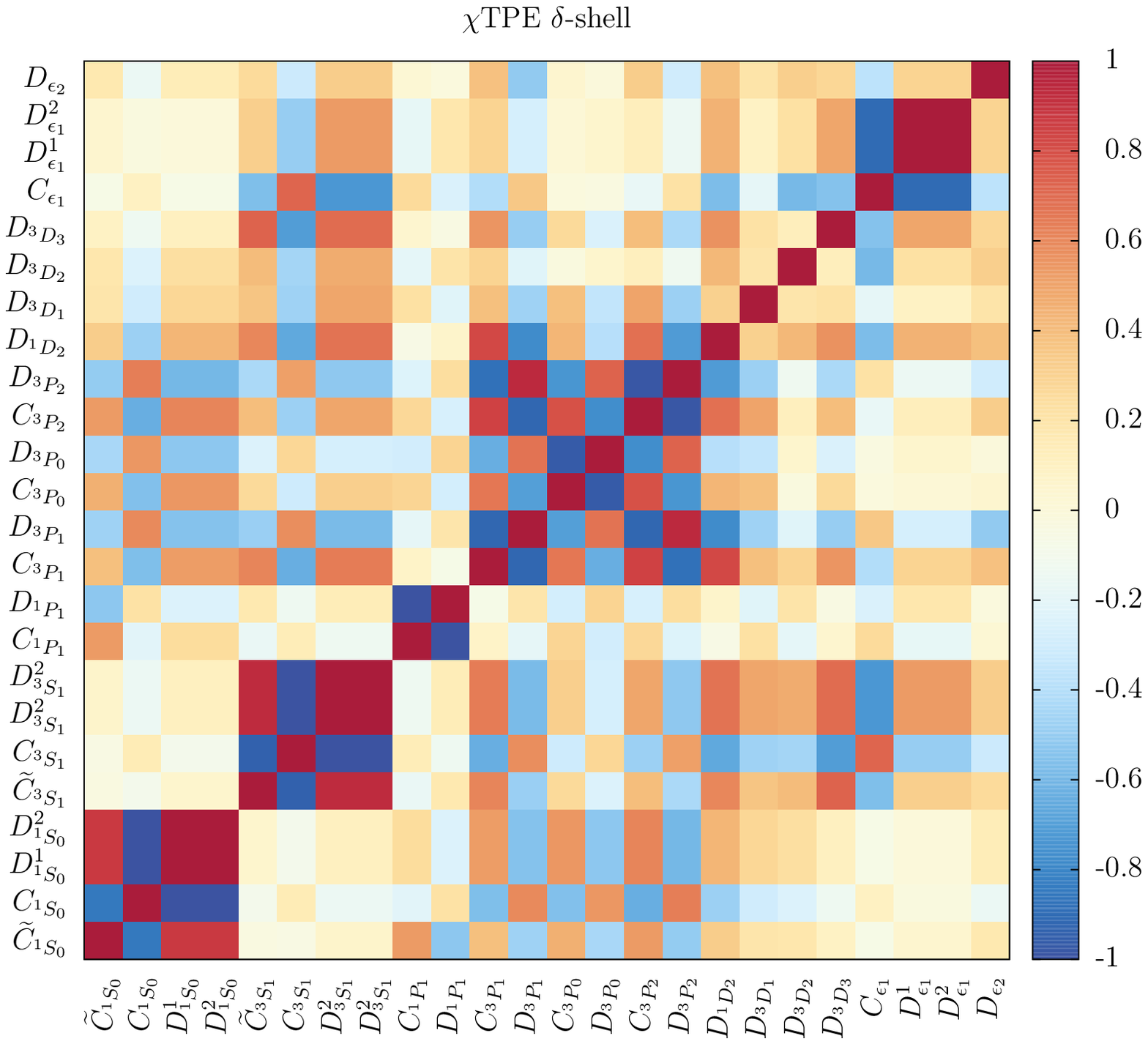,width=6cm}  
\end{center}
\caption{
(Color online) Same as Fig~\ref{fig:corr-skyrme} 
but for the short distance contribution to the 24 partial wave
  counterterms including up to ${\cal O} (p^4)$ }
\label{fig:corr-ct}
\end{figure}

The potential in momentum space can be written 
in the partial wave basis as 
\begin{eqnarray}
v^{JS}_{l',l} (p',p) =(4\pi)^2 \int_0^\infty \, dr\, r^2 \, j_{l'}(p'r)
j_{l}(pr) V_{l' l}^{JS}(r) \,
\end{eqnarray}
Using the Bessel function expansion for small argument $ j_l(x) =
x^l/(2l+1)!!  [1 - x^2/2(2l+3)+ \dots ] $ we get a low momentum
expansion of the potential matrix elements. 
We keep up to total order
${\cal O} (p^4, p'^4 , p^2 p'^2)$ corresponding to $S$-, $P$- and
$D$-waves as well as S-D and P-F mixing parameters,
\begin{eqnarray}
v_{00}^{JS}(p',p) 
&=& 
 \widetilde{C}_{00}^{JS} 
+ C_{00}^{JS}(p^2+p'^2) 
+ D^1_{00}{}^{JS} (p^4+p'^4) 
+ D^2_{00}{}^{JS} p^2 p'^ 2 
+ \cdots 
\nonumber \\ 
v_{11}^{JS}(p',p) 
&=& 
p p' C_{11}^{JS} 
+ p p' (p^2+p'^2) D_{11}^{JS} 
+ \cdots 
\nonumber \\
v_{22}^{JS}(p',p) 
&=& 
p^2 p'{}^2 D_{22}^{JS} 
+ \cdots 
\nonumber   \\ 
v_{20}^{JS}(p',p) 
&=& p'{}^2 C_{20}^{JS} 
+ p'{}^2 p^ 2  D^1_{20}{}^{JS} + p'{}^4 D^2_{20}{}^{JS} 
+ \dots \nonumber   \\
v_{31}^{JS}(p',p) 
&=& 
p'{}^3 p D_{31}^{JS} 
+ \cdots 
\label{eq:Cts}
\end{eqnarray}
We use the spectroscopic notation and normalization of
Ref.~\cite{Epelbaum:2004fk}. We will call the coefficients in the
expansion counterterms~\footnote{Properly speaking the name is
  justified when the potential $v (p',p)$ is used to solve the problem
  in a restricted Hilbert space $p,p' \le \Lambda$ which means in
  particular fitting scattering data up to CM momentum $p \le
  \Lambda$~\cite{Arriola:2013era}. Only under these conditions is a
  truly universal behaviour of the counterterms guaranteed for
  $\Lambda \lesssim 1/a$.}. Our numerical results are shown in
table~\ref{tab:LECS} for OPE-DS~\cite{Perez:2013jpa} and
$\chi$TPE-DS~\cite{Perez:2013oba}.  In both cases it is interesting to
separate the short distance contribution from the explicit potential
tail containing the pion contributions.  As we see, while the separate
contributions greatly differ from one region to another, there is a
high degree of universality for the full integral, which accounts for
the integrated strength of the interaction.  The correlation matrices
are presented in Fig.~\ref{fig:corr-ct} for the inner $r < r_c$
contributions. The short distance contribution to the counterterms in
the partial wave basis are largely independent, although the $\chi$TPE
enhances correlations as compared to OPE, see
Fig.~\ref{fig:corr-ct}. This correlation information is actually very
useful to find a flexible minimization path in both
OPE-DS~\cite{Perez:2013jpa} and $\chi$TPE-DS~\cite{Perez:2013oba}
potentials and complies to the same correlation pattern obtained in
our previous work~\cite{Perez:2014yla}. In any case it is easier to
take the short distance parameters as primary fitting quantities and
the counterterms as derived ones.

Note that a direct numerical comparison with momentum space
treatments~\cite{Epelbaum:2004fk,Ekstrom:2013kea} is tricky since the
particular choice of splitting the pionic and short distance
interaction is done differently.  This can be seen in the numerical
values given in ~\cite{Epelbaum:2004fk} and \cite{Ekstrom:2013kea}
where the same convention and chiral potential is used but different
fitting ranges are considered. In addition we find that there are
systematic differences within our quoted statistical
uncertainties, table~\ref{tab:LECS} (no statistical errors are quoted
in~\cite{Epelbaum:2004fk} and \cite{Ekstrom:2013kea}).  In all we show
24-low energy constants in the np case, but note that they stem from
42 np parameters in the OPE case and 27 in the $\chi$TPE case, so we
could have taken these low energy constants themselves as fitting
parameters. 

\begin{table}[ht]
 \caption{\label{tab:LECS} Potential integrals in different partial
   waves.  We separate the contribution from the delta-shells short
   range parameters (corresponding to $r < r_c$) and the potential
   tail (corresponding to $r> r_c$) both for OPE-DS ($r_c=3 {\rm fm}$)~\cite{Perez:2013jpa} and $\chi$TPE-DS ($r_c=1.8 {\rm fm}$)~\cite{Perez:2013oba}. Units
   are: $\widetilde{C}$'s are in $10^4 {\rm GeV}^{-2}$, $C$'s are in
   $10^4 {\rm GeV}^{-4}$ and $D$'s are in $10^4 {\rm GeV}^{-6}$.}
\begin{center}
 \begin{tabular}{c c c c  c c c }
\hline
                       & \multicolumn{3}{c}{OPE-DS($r_c=3 {\rm fm}$)} &  \multicolumn{3}{c}{$\chi$TPE-DS($r_c=1.8 {\rm fm}$)} \\[-5pt]
                     &   $ r < r_c $ & $ r > r_c $ & Full &     $ r < r_c $ & $ r > r_c $ & Full \\              
\hline                     
$\widetilde{C}_{^1S_0}$   &  -0.120(1)  &   -0.021      &   -0.141(1) &  -0.071(2)  &   -0.0641(5)  &   -0.135(2)  \\[-5pt]
            $C_{^1S_0}$  &   1.83(2)   &    2.34       &    4.17(2)  &  0.74(2)   &    3.384(9)   &    4.12(2)    \\[-5pt]
            $D_{^1S_0}^1$   &-39.2(11)   & -409.6        & -448.8(11)  & -6.1(3)    & -437.6(4)     & -443.7(5)      \\[-5pt]
            $D_{^1S_0}^2$ &-11.8(3)    & -122.8        & -134.6(3)   & -1.83(9)   & -131.3(1)     & -133.1(1)     \\[-5pt]
$\widetilde{C}_{^3S_1}$  &  -0.041(2)  &   -0.023      &   -0.064(2) &   0.015(2)  &   -0.0532(5)  &   -0.038(1)    \\[-5pt]
            $C_{^3S_1}$  &   1.18(1)   &    2.61       &    3.79(1)  &   0.23(1)   &    3.33(1)    &    3.55(1)    \\[-5pt]
            $D_{^3S_1}^1$ &-24.3(3)    & -486.4        & -510.7(3)   &  -3.8(2)    & -500.9(4)     & -504.7(4)   \\[-5pt]
            $D_{^3S_1}^2$ & -7.3(1)    & -145.9        & -153.2(1)   &   -1.15(6)   & -150.3(1)     & -151.4(1)    \\[-5pt]
            $C_{^1P_1}$   &  1.21(2)   &    5.23       &     6.44(2) &   0.551(6)  &    5.99(1)    &    6.54(1)    \\[-5pt]
            $D_{^1P_1}$  & -11.2(2)    & -583.7        &  -594.9(2)  &  -2.04(2)   & -590.1(2)     & -592.1(2)      \\[-5pt]
            $C_{^3P_1}$  &   1.151(3)  &    2.587      &     3.738(2) &   0.663(5)  &    2.997(5)   &    3.659(3)   \\[-5pt]
            $D_{^3P_1}$  &  -9.44(5)   & -243.85       &  -253.29(5)  &    -2.55(5)   & -247.3(1)     & -249.8(2)  \\[-5pt]  
            $C_{^3P_0}$  &  -1.298(8)  &   -3.613      &    -4.911(8) & 0.037(5)  &   -4.918(6)   &   -4.882(5)     \\[-5pt]
            $D_{^3P_0}$  &  23.1(2)    &  323.9        &   347.0(2)   &   1.17(5)   &  342.5(2)     &  343.6(2)        \\[-5pt]
            $C_{^3P_2}$  &  -0.552(2)  &    0.107      &    -0.445(2) &    -0.234(5)  &   -0.201(5)   &   -0.434(3)     \\[-5pt]
            $D_{^3P_2}$  &   6.14(7)   &  -16.76       &   -10.62(7)  &  1.18(4)   &  -10.8(2)     &   -9.7(2)     \\[-5pt]
            $D_{^1D_2}$  &  -5.38(3)   &  -65.54       &   -70.92(3)  &  -0.65(3)   &  -70.01(6)    &  -70.66(6)    \\[-5pt]
            $D_{^3D_2}$  & -17.0(2)    & -350.8        &  -367.8(2)   &   -1.87(3)   & -362.52(6)    & -364.39(7)    \\[-5pt]
            $D_{^3D_1}$  &  10.7(2)    &  195.1        &   205.8(2)   &     1.51(4)   &  202.74(7)    &  204.25(7)    \\[-5pt]
            $D_{^3D_3}$  &   0.41(1)   &    0.14       &     0.55(1)  &       0.54(1)   &    0.33(6)    &    0.87(6)      \\[-5pt]
       $C_{\epsilon_1}$   & -2.38(2)   &   -5.98       &    -8.36(2)  &   -0.906(6)  &   -7.594(4)   &   -8.500(4)     \\[-5pt]
       $D_{\epsilon_1}^1$ & 47.7(6)    &  964.9        &  1012.6(6)   &  6.73(7)   &  998.8(1)     & 1005.5(1)     \\[-5pt]
       $D_{\epsilon_1}^2$ & 20.4(3)    &  413.6        &   434.0(3)   &   2.88(3)   &  428.06(5)    &  430.94(4)    \\[-5pt]
       $D_{\epsilon_2}$   &  4.72(4)   &   79.46       &    84.18(4) &     0.709(6)  &   82.58(1)    &   83.29(1)  \\
\hline
 \end{tabular}
\end{center}
\end{table}

\subsection{Discussion}

The universality and lack of correlation among fitting parameters are
good features of a least squares minimization procedure.  Our results
above address the issue and a clear picture emerges.  From an EFT
point of
view~\cite{Weinberg:1990rz,Kaiser:1997mw,Entem:2003ft,Epelbaum:2004fk,Machleidt:2011zz}
the partial wave parameters are nothing but the needed counterterms at
a given renormalization scale $\Lambda$, as they encode the integrated
out short range information of the interaction.  We can also see that
the correlation matrix among the different counterterms indicates to
what extent are they intertwined. This is important since in EFT
counterterms are assumed to be independent variables. Discovering
correlations among them means that some extra symmetry or condition
has been overlooked. It is remarkable that at $\Lambda\sim 200 {\rm
  MeV}$ a long distance symmetry such as the spin-isospin SU(4)-Wigner
invariance emerges from the NN data and simultaneously is expected
from a large $N_c$
approach~\cite{Kaplan:1996rk,CalleCordon:2008cz,CalleCordon:2009ps}.
This pattern does not show up in Fig.~\ref{fig:corr-ct}
corresponding to $\Lambda=400 {\rm MeV}$ but without restricting the
full Hilbert space for off-shell momenta $p,p' \le \Lambda$. This
suggests a correlation analysis closer in spirit to the BD-SRG
approach~\cite{Anderson:2008mu,Arriola:2013era}.

\section{Conclusions}

There is currently an unbalanced accuracy between theory and
experiment in Nuclear Physics. This is due not only to the complexity
of the nuclear many body problem but also because the fundamental
nucleon-nucleon interaction is not known precisely in the mid-range
region which is the relevant one for nuclear binding. On top of this
the precise level of inaccuracy of the nuclear force is rarely
estimated. This is a major handicap for the assessment of theoretical
uncertainties when going from the two body problem to the many body
situation. At present, the most accurate source of information of the
two body interaction is the large collection of about $8000$ np and pp
scattering data, which by themselves are subjected to experimental
uncertainties. The disentanglement of systematic and statistical
errors, while challenging in itself, can be simplified by selecting
scattering data which are not only mutually consistent but also true
normal fluctuations with respect to the most likely phenomenological
interaction. This desirable scenario, while is not at all guaranteed,
does actually happen in our analysis and furnishes the necessary
requirements for a sound statistical error propagation. We have
discussed with some detail the statistical considerations to test the
normality of residuals with a stringent and recently proposed tail
sensitive scheme.  We also provide a straightforward recipe to carry
out such a useful test for any least squares fit. As a particular but
insightful application, we have also profited from the applicability
of statistical error and correlation analysis to quantify the
uncertainty of effective interactions defined as volume integrals of
the potential. As we have noted, effective interactions depend on the
resolution scale determined by the shortest de Broglie wavelength
involved. We have focused and restricted on a relatively small
distance scale determined by the pion production threshold, where the
repulsive short distance features become relevant. As preliminary
calculations show the scale dependence could actually be tuned to the
relevant de Broglie wavelength dominating the NN interaction in finite
nuclei.  In any case a certain degree of universality emerges from our
analysis, and reinforces the view that effective interaction
parameters have desirable features of fitting parameters, namely small
statistical correlations and scheme independence.

\bigskip
This work is supported by Spanish DGI (grant FIS2011-24149)
and Junta de Andaluc\1a (grant FQM225). R.N.P. is supported by a
Mexican CONACYT grant.

\appendix


\section{Tables on Tail Sensitive Test Statistic}


\begin{table*}
 \caption{\label{tab:TStestcritical} Critical values of $T_{\rm c}$
   for the Tail Sensitive normality test as a function of the sample
   size $N$ at different levels of significance $\alpha$. The critical
   values were obtained using MonteCarlo simulations taking 500000
   samples for each value of $N$ between $1$ and $50$.}
 \begin{tabular*}{\columnwidth}{@{\extracolsep{\fill}} l l l l l l}
 $ N \backslash \alpha$  &   0.01        &   0.02        &   0.05        &   0.10        &   0.20       \\
    \hline 
  1  &   0.01001773  &   0.02009725  &   0.05004725  &   0.10012406  &   0.19996581 \\[-8pt]
  2  &   0.00513134  &   0.01036312  &   0.02648217  &   0.05448341  &   0.11389617 \\[-8pt]
  3  &   0.00351999  &   0.00718572  &   0.01862308  &   0.03892356  &   0.08284280 \\[-8pt]
  4  &   0.00274579  &   0.00559971  &   0.01464807  &   0.03088211  &   0.06658627 \\[-8pt]
  5  &   0.00225585  &   0.00466067  &   0.01228834  &   0.02606941  &   0.05682341 \\[-8pt]
  6  &   0.00194960  &   0.00403137  &   0.01064535  &   0.02273282  &   0.04997837 \\[-8pt]
  7  &   0.00171087  &   0.00355567  &   0.00950508  &   0.02041778  &   0.04507288 \\[-8pt]
  8  &   0.00153800  &   0.00320521  &   0.00862923  &   0.01856308  &   0.04123497 \\[-8pt]
  9  &   0.00142940  &   0.00296436  &   0.00796641  &   0.01716920  &   0.03823687 \\[-8pt]
 10  &   0.00130562  &   0.00272923  &   0.00739200  &   0.01599717  &   0.03579932 \\[-8pt]
 11  &   0.00121215  &   0.00255493  &   0.00693178  &   0.01504779  &   0.03371454 \\[-8pt]
 12  &   0.00115210  &   0.00241817  &   0.00654682  &   0.01423879  &   0.03203655 \\[-8pt]
 13  &   0.00108994  &   0.00229341  &   0.00620929  &   0.01353055  &   0.03054902 \\[-8pt]
 14  &   0.00103473  &   0.00217373  &   0.00592355  &   0.01295297  &   0.02922865 \\[-8pt]
 15  &   0.00098603  &   0.00208026  &   0.00568029  &   0.01243297  &   0.02814438 \\[-8pt]
 16  &   0.00094095  &   0.00199030  &   0.00545322  &   0.01193342  &   0.02711705 \\[-8pt]
 17  &   0.00091111  &   0.00192155  &   0.00524212  &   0.01153270  &   0.02622548 \\[-8pt]
 18  &   0.00087656  &   0.00185463  &   0.00508330  &   0.01116882  &   0.02539659 \\[-8pt]
 19  &   0.00084824  &   0.00178377  &   0.00490132  &   0.01079536  &   0.02462677 \\[-8pt]
 20  &   0.00081993  &   0.00173589  &   0.00475800  &   0.01049558  &   0.02399643 \\[-8pt]
 21  &   0.00079345  &   0.00168404  &   0.00464703  &   0.01023746  &   0.02338279 \\[-8pt]
 22  &   0.00077672  &   0.00164040  &   0.00451678  &   0.00995987  &   0.02278223 \\[-8pt]
 23  &   0.00075279  &   0.00159405  &   0.00441474  &   0.00974473  &   0.02233426 \\[-8pt]
 24  &   0.00073934  &   0.00156243  &   0.00430510  &   0.00953482  &   0.02183490 \\[-8pt]
 25  &   0.00071677  &   0.00151937  &   0.00421339  &   0.00930246  &   0.02137357 \\[-8pt]
 26  &   0.00069494  &   0.00148552  &   0.00412071  &   0.00912411  &   0.02092654 \\[-8pt]
 27  &   0.00068503  &   0.00145519  &   0.00401387  &   0.00891337  &   0.02055165 \\[-8pt]
 28  &   0.00066741  &   0.00141934  &   0.00394817  &   0.00875921  &   0.02018670 \\[-8pt]
 29  &   0.00065925  &   0.00139673  &   0.00387957  &   0.00861749  &   0.01982410 \\[-8pt]
 30  &   0.00064347  &   0.00137429  &   0.00380204  &   0.00845258  &   0.01949674 \\[-8pt]
 31  &   0.00063372  &   0.00134370  &   0.00374100  &   0.00830829  &   0.01920333 \\[-8pt]
 32  &   0.00062434  &   0.00133204  &   0.00368650  &   0.00819084  &   0.01892001 \\[-8pt]
 33  &   0.00061108  &   0.00130865  &   0.00362774  &   0.00805833  &   0.01865113 \\[-8pt]
 34  &   0.00060333  &   0.00128911  &   0.00357679  &   0.00796122  &   0.01842678 \\[-8pt]
 35  &   0.00059576  &   0.00126498  &   0.00352836  &   0.00784902  &   0.01816494 \\[-8pt]
 36  &   0.00058123  &   0.00124395  &   0.00347131  &   0.00772213  &   0.01790285 \\[-8pt]
 37  &   0.00057221  &   0.00122973  &   0.00342757  &   0.00764118  &   0.01770791 \\[-8pt]
 38  &   0.00056670  &   0.00120748  &   0.00336587  &   0.00752593  &   0.01744823 \\[-8pt]
 39  &   0.00056443  &   0.00119716  &   0.00334196  &   0.00744526  &   0.01726910 \\[-8pt]
 40  &   0.00055411  &   0.00118668  &   0.00330584  &   0.00735387  &   0.01704568 \\[-8pt]
 41  &   0.00054087  &   0.00116426  &   0.00324657  &   0.00726325  &   0.01688035 \\[-8pt]
 42  &   0.00053338  &   0.00114686  &   0.00321383  &   0.00717868  &   0.01668878 \\[-8pt]
 43  &   0.00053060  &   0.00113421  &   0.00317425  &   0.00710159  &   0.01651373 \\[-8pt]
 44  &   0.00052550  &   0.00113043  &   0.00315774  &   0.00702840  &   0.01634427 \\[-8pt]
 45  &   0.00051907  &   0.00110859  &   0.00311188  &   0.00695394  &   0.01619440 \\[-8pt]
 46  &   0.00051066  &   0.00109194  &   0.00307711  &   0.00689179  &   0.01605360 \\[-8pt]
 47  &   0.00050927  &   0.00109077  &   0.00304886  &   0.00681100  &   0.01583361 \\[-8pt]
 48  &   0.00050123  &   0.00107538  &   0.00301819  &   0.00675336  &   0.01569550 \\[-8pt]
 49  &   0.00049858  &   0.00107373  &   0.00298081  &   0.00668439  &   0.01557817 \\[-8pt]
 50  &   0.00049267  &   0.00105755  &   0.00296376  &   0.00663403  &   0.01544530 
 \end{tabular*}
\end{table*}

\begin{figure}[ht]
\begin{center}
\epsfig{figure=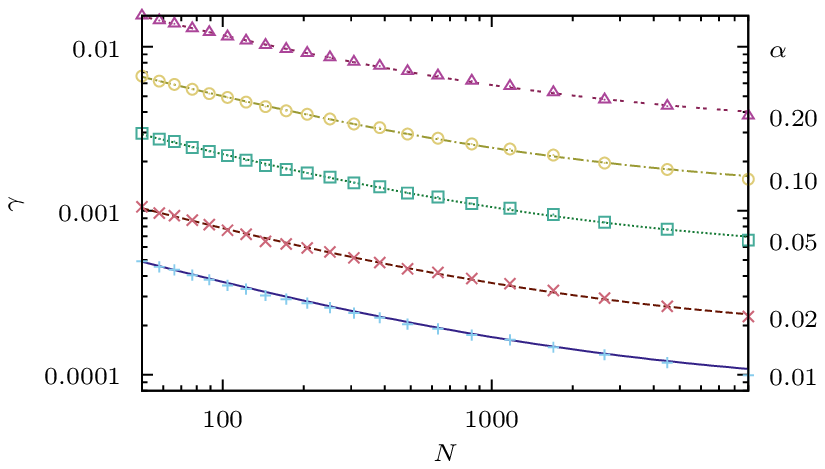,width=7cm}  
\end{center}
\caption{(Color online) Critical values for the Tail Sensitive
  normality test for large values of N at different levels of
  significance $\alpha$. The points where obtained via MonteCarlo
  simulations taking 500000 samples for each value of $N$. The curves
  correspond to the parameterization $T_{\rm c} = a/\sqrt{N} + b$; the
  fitted values for $a$ and $b$ at each level of significance are
  given in table~\ref{tab:TSparameters}.}
\label{Fig:largeNFit}
\end{figure}

\begin{table}
 \caption{\label{tab:TSparameters} Large $N$ parameterization of the critical values for the
 Tail Sensitive normality test $T_{\rm c} = a/\sqrt{N} + b$ at different levels of significance. This parameterization 
 is appropriate for $50 < N \leq 9000$}
\begin{center}
 \begin{tabular}{ c c c  }
   $\alpha$&   a       & b         \\
    \hline 
    0.01   & 0.0029065 & 0.0000728 \\ [-8pt]
    0.02   & 0.0061401 & 0.0001690 \\[-8pt]
    0.05   & 0.0169933 & 0.0005161 \\[-8pt]
    0.10   & 0.0377146 & 0.0012314 \\[-8pt]
    0.20   & 0.0866187 & 0.0031098 \\
\hline 
 \end{tabular}
\end{center}
\end{table}

The critical values in table~\ref{tab:TStestcritical} were obtained by
taking $M=500000$ random samples with size $N$ of uniformly
distributed data. The rather high number of samples was necessary to
reduced the statistical fluctuations enough so that if $N_1 < N_2 \leq
50$ then $T_{1c,\rm TS} > T_{2c,\rm TS}$. One of the appeals of the
Kolmogorov-Smirnov test is the simple $\sim 1/\sqrt{N}$
parameterization of the critical values for large $N$.  In
Ref.~\cite{Aldor2013} the authors proposed a $\sim
\log{(\log{(N)})}/\sqrt{N}$ limiting behavior, but the parameters
values were not determined. We performed new Monte-Carlo simulations,
maintaining the number of samples, to obtain the reliable critical
values for several values of $N$ between $50$ and $9000$ and find a
valid parameterization in this range of $N$.  We tried the proposed
parameterization by Aldor-Noiman et al. but found a better fit to our
results, shown in Fig. \ref{Fig:largeNFit}, by using
\begin{equation}
 T_{c,\rm TS} = \frac{a}{\sqrt{N}} + b.
\end{equation}
The fitted parameters for different levels of significance are
presented in table~\ref{tab:TSparameters}. It is important to mention
that this parameterization is valid in the $50 < N \leq 9000$ range and
extrapolations to $N > 9000$ may not be correct




\vskip.2cm 


\end{document}